\documentclass[preprint2]{aastex}

\usepackage{amsmath}
\usepackage{graphicx,multicol}
\usepackage{times}
\def\etal{\it et~al.}

\shorttitle{Search for Off-pulse emission in Pulsars}
\shortauthors{Basu, Mitra \& Melikidze}

\begin{document}

\title{Search for Off-pulse emission in Long Period Pulsars}
\author{Rahul Basu\altaffilmark{1}, Dipanjan Mitra\altaffilmark{2,3}, George I. Melikidze\altaffilmark{3,4}} 

\altaffiltext{1}{Inter-University Centre for Astronomy and Astrophysics, Pune, 411007, India; rahulbasu.astro@gmail.com}
\altaffiltext{2}{National Centre for Radio Astrophysics, Tata Institute of Fundamental Research, Pune 411007, India.} 
\altaffiltext{3}{Janusz Gil Institute of Astronomy, University of Zielona G\'ora, ul. Szafrana 2, 65-516 Zielona G\'ora, Poland.}
\altaffiltext{4}{Evgeni Kharadze Georgian National Astrophysical Observatory, 0301 Abastumani, Georgia.}

\begin{abstract}
We have revisited the problem of off-pulse emission in pulsars, where detailed 
search for the presence of low level radio emission outside the pulse window is
carried out. The presence of off-pulse emission was earlier reported in two 
long period pulsars, PSR B0525+21 and B2046--16 at frequencies below 1 GHz 
using the Giant Meterwave Radio Telescope (GMRT). However, subsequent studies 
did not detect off-pulse emission from these pulsars at higher radio 
frequencies ($>$1 GHz). We have carefully inspected the analysis scheme used in
the earlier detections and found an anomaly with data editing routines used, 
which resulted in leakage of signal from the on-pulse to the off-pulse region. 
We show that the earlier detections from PSR B0525+21 and B2046--16 were a 
result of this leakage. The above analysis scheme has been modified and 
offline-gating has been used to search for off-pulse emission in 21 long period
pulsars ($P > 1.2$ sec) at different observing frequencies of GMRT. The 
presence of low level off-pulse emission of peak flux 0.5 mJy was detected in 
the brightest pulsar in this list PSR 0B0628--28, with off-pulse to average 
pulsar flux ratio of 0.25\%. We suggest that coherent radio emission resulting 
due to cyclotron resonance near the light cylinder can be a possible source for
the off-pulse emission in this pulsar.
\end{abstract}


\section{Introduction}
\noindent
The radio emission from pulsars originates around heights of $\sim$ 500 km from
the stellar surface, which is less than 10\% of the light cylinder radius 
\citep{kij98,mit02,kij03,mit04,krz07,mit17}. As a result the observed radio 
emission is primarily seen as narrow pulses, the main pulse, which usually 
occupy a small fraction ($\sim$ 10\%) of the period. In rare cases, when the 
rotation axis is either very close to the dipolar magnetic axis or they are 
orthogonal to each other, the main pulse is either seen over a large fraction 
of the pulsar period or inter-pulse emission from the opposite pole is visible.
The other pulsed emission originating outside the main pulse include the 
pre/post-cursor emission \citep{bas15}, whose origin in the outflowing 
relativistic plasma is particularly challenging to understand. 

There has been a number of works dedicated to search for and study the presence
of un-pulsed emission from pulsars \citep{per85,bar85,str90,han93,sta99,bas11,
bas12,mar19}. In these works primarily interferometric techniques were 
employed to detect low level emission in the profile baseline region 
(off-pulse), where the main pulse (on-pulse) was masked or `gated'. A 
scintillation based search for off-pulse emission has also been proposed by 
\citet{rav18}. The motivation for off-pulse emission studies either involve 
investigation of extended nebulae around pulsars \citep{wei74,gae98,gae00,
dzi18,rua20} or the presence of coherent emission higher up the pulsar 
magnetosphere \citep{bas13}. Other motivation for off-pulse studies 
involves the recent search for axions, which are a dark matter candidate, 
towards pulsars \citep{dar20a,dar20b}. The spectral line searches for these 
particles will have improved detection sensitivity in the off-pulse window and 
reduce the effects of scintillation on continuum baselines.

The presence of off-pulse emission from two long period pulsars B0525+21 ($P=$ 
3.746 sec) and B2046--16 ($P=$ 1.961 sec) was reported by \citet{bas11,bas12} 
using the 325 MHz and 610 MHz frequency bands of the Giant Meterwave Radio 
Telescope (GMRT). These studies used the technique of offline-gating, where 
high time resolution interferometric data was recorded and subsequently 
divided into on-pulse and off-pulse parts which were imaged separately. The 
on-pulse and off-pulse emission showed contemporaneous variations in intensity 
due to interstellar scintillation which suggested the off-pulse emission to 
emerge close to the on-pulse and thereby have a magnetospheric origin. 
\citet{mar19} has carried out high spatial resolution observations of the two 
pulsars B0525+21 and B2046--16 using the European Very Large Baseline Network 
(EVN) at 1.39 GHz frequencies, with phase-resolved visibilities across the 
pulsar period. They did not detect any significant off-pulse emission in either
pulsar. This prompted us to re-examine the analysis scheme used for the 
previous detections of off-pulse emission in \citet{bas11,bas12}. It was found 
that the data editing technique used in these studies led to leakage of signal 
from the on-pulse to the off-pulse window. As a result spurious emission 
appeared in the off-pulse images in the location of the pulsar. We have 
modified the analysis scheme to remove the source of leakage and extended the 
offline-gating studies to a larger sample of long period pulsars. In section 
\ref{sec:leak} we describe the source of leakage signal from the on-pulse to 
the off-pulse window and the corrective measures applied to the analysis 
scheme. In section \ref{sec:obs} we present the details of the observations of 
21 long period pulsars and the offline-gating used to probe the presence of 
off-pulse emission in them. Section \ref{sec:result} shows the results where 
low level emission is seen in the off-pulse window of the brightest pulsar 
(B0628--28) in the sample, and we discuss the implications of the results in 
section \ref{sec:disc}.

\begin{figure*}
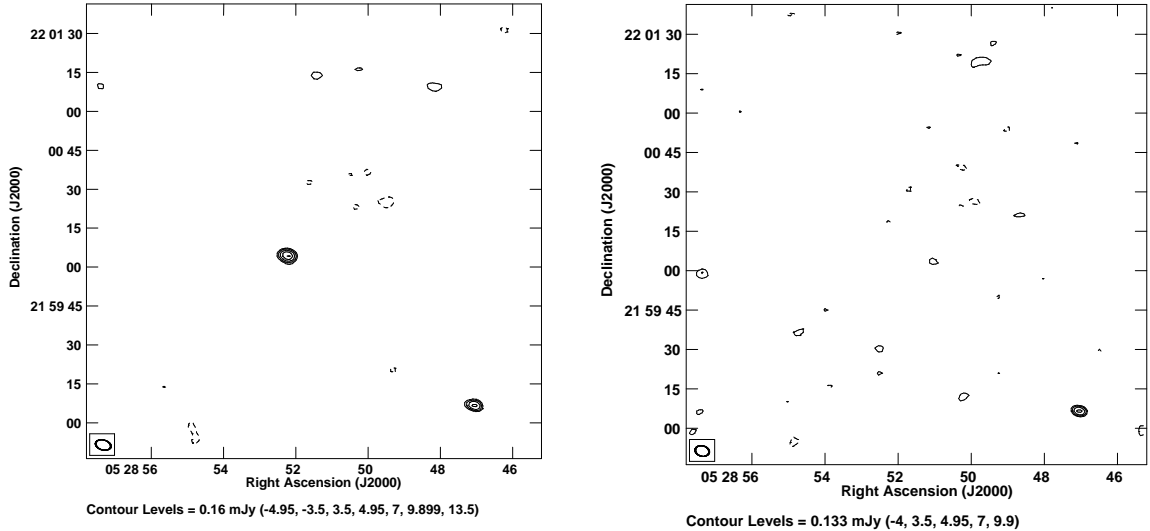

\centering
\begin{tabular}{@{}cr@{}}
{\mbox{\includegraphics[scale=0.39,angle=0.]{B0525F_NF_22jul.ps}}} &
{\mbox{\includegraphics[scale=0.43,angle=0.]{B0525F_22jul.ps}}} \\
\end{tabular}
\caption{The figure shows the effect of leakage on the off-pulse image of PSR 
B0525+21, with observations carried out at frequencies of 610 MHz on 22 July, 
2011. The images represent the total intensity contours centered at the 
location of the pulsar. The right-hand panel shows a point source at the 
location of the pulsar which arises due to leakage of signal from the on-pulse 
to the off-pulse window. The left-hand panel shows the off-pulse map where the 
leakage has been corrected and no off-pulse emission is seen. The resolution in
each image is shown in the bottom left box, while an unrelated nearby point 
source is also visible near the bottom right corner.}
\label{fig:leak}
\end{figure*}

\section{Investigating Previous Detection of Off-pulse Emission}\label{sec:leak}

\noindent
\citet{bas11} developed the offline-gating technique to investigate the 
presence of off-pulse emission using the Giant Meterwave Radio Telescope 
\citep[GMRT,][]{swa91}. This requires high time resolution interferometric 
observations such that the pulsar period can be divided into sufficient number 
of bins to separate the on and off-pulse regions and image them individually. 
Before the gating process an automatic editing software was used to remove 
Radio Frequency Interference (RFI) by \citet{bas11,bas12}. It was found that 
the data editing software caused leakage of on-pulse signal into the off-pulse 
region. This resulted in spurious detection of off-pulse emission in these 
earlier works. An example of this leakage leading to off-pulse emission in PSR 
B0525+21 is shown in Fig. \ref{fig:leak}. The left-hand panel corresponds to an
image of the off-pulse region in the presence of leakage and shows a prominent 
point source at the location of the pulsar. The right-hand panel shows the 
equivalent image without leakage and no off-pulse emission is detected.

Astronomical observations are generally recorded in Flexible Image Transport 
System (FITS) format. In case of radio interferometry the FITS comprises of 
Header Data Units (HDU) which records observing details, like the time of 
observation, the baseline pair identifier, source identifier, etc. The 
visibility measurements by the interferometer is recored as a binary sequence 
following the HDU. In case of standard GMRT observations, spanning a certain 
frequency range separated by equispaced channels and more than one 
polarization, the visibility for each time-baseline-channel-polarization unit 
is stored as a set of three numbers, real, imaginary measurements and a 
specific weighting for each measurement. When the weight is `1' the 
visibilities are read unchanged while they are ignored when the weight is `0', 
and any other value scales them accordingly. A continuous series of such three 
element sets are recored for all measurements. The manipulation of FITS files 
can be carried out using standard softwares like the cfitsio package in the c 
programming language.
 
In the data editing process a subset of the entire observation is initially 
read in a buffer array and the statistics is estimated. If any particular 
visibility is found to be a statistical outlier, it is considered to be  
affected by RFI and not used in the subsequent analysis by changing their 
weights to `0' in the buffer. Finally, the visibilities identified as outliers 
are copied from the buffer to the FITS dataset, where the exchange happens only
if any statistical outliers are found. However, we have found a bug in the 
editing software where the data is copied into a different location in the 
initial FITS file from the buffer. This results in leakage of the on-pulse 
signal into the off-pulse region where they are overwritten. This leakage has 
caused the detection of off-pulse emission reported by \citet{bas11,bas12}, as 
shown in Fig. \ref{fig:leak} left-hand panel, which are spurious detections. 
Around 5-10\% of the observing durations were statistical outliers, which is 
comparable to the ratio between the on-pulse and off-pulse flux levels in these
studies. The off-pulse followed the scintillation behaviour of the on-pulse 
signal as they were low level on-pulse emission. We have modified our analysis 
scheme to ensure all data editing is carried out after the gating stage, where 
the original file is separated into two, corresponding to the on-pulse and 
off-pulse windows. The data editing softwares have been updated to ensure that 
the location of all visibilities are correctly identified in both the buffer 
and the FITS datasets. An example of the result of new analysis is show in Fig.
\ref{fig:leak} right-hand panel, where no clear emission at the location of the
pulsar is visible in the off-pulse image. The analysis was extended to the 
previous observations of the pulsars B0525+21 and B2045--16 reported in 
\cite{bas11,bas12}, where no clear off-pulse emission was seen. The remaining 
measurements, including the on-pulse flux and the noise levels in the on-pulse 
and off-pulse images were unchanged.

\begin{figure*}
\centering
\includegraphics[scale=0.7,angle=0.]{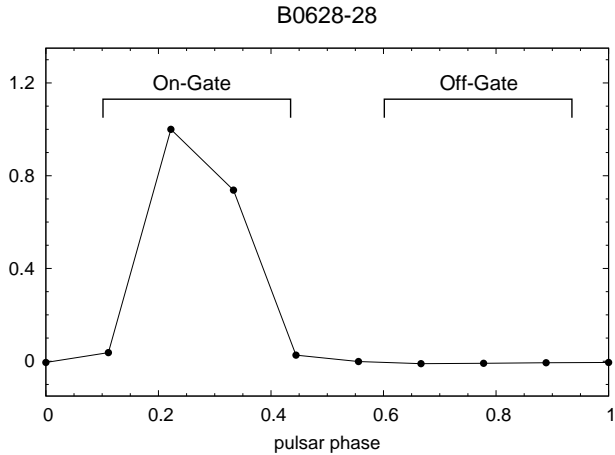} 
\caption{The folded profile of PSR B0628--28 from high time resolution 
interferometric observations. The on-pulse and off-pulse windows for the images
are shown in each profile.}
\label{fig:fold}
\end{figure*}

\section{Observation \& Analysis}\label{sec:obs}
We have carried out extended observations to search for off-pulse emission in 
pulsars using GMRT, which consists of 30 separate antennas arranged in an 
Y-shaped array with maximum distance of 27 kilometers between antenna pairs. 
The observations were carried out using the GMRT software correlator 
\citep[GSB,][]{roy10}, which is currently decommissioned and replaced with a 
wide-band backend. Interferometric observations generally measure visibilities, 
which are correlated signals from antenna pairs, after averaging over several 
seconds and hence are not suitable for most pulsar studies, where the radio 
emission is seen as narrow bursts of emission repeating at regular periods 
ranging from few milliseconds to several seconds. A special mode of the GSB 
allowed observations with high time resolutions of 128/256 milliseconds that 
were used for these studies. It is also possible to record the self data, which
is the auto-correlated signals from each antenna, measuring the absolute 
intensity of the incoming signals. Offline-gating technique was used for these 
studies, where self-data from pulsars were initially folded at their rotating 
period to obtain an average profile with well defined on-pulse and off-pulse 
regions (see Fig. \ref{fig:fold}). Each recorded time was subsequently 
assigned a phase related to the profile and the observation was separated into 
two parts, corresponding to the on-pulse and off-pulse regions \citep[see][for 
additional details]{bas11}. Standard imaging techniques using classic AIPS 
package was utilized to image the on-pulse and off-pulse parts and search for 
off-pulse emission in the location of the on-pulse source. The observations 
were conducted between 2013 and 2014, when the GSB was still operational, at 
two frequencies centered around 610 MHz and 1280 MHz with 33 MHz bandwidth. 

The 128 millisecond time resolution of GMRT interferometric observations 
limited the number of pulsars suitable for exploring the presence of off-pulse 
emission. At least 8-9 phase-bins in the profile window are essential to 
clearly separate the On-pulse and off-pulse regions of the profile \citep[see 
discussion in][]{bas11,bas12}. As a result long period pulsars with $P>$ 1.2 
seconds were considered for this work. In addition, we considered nearby bright
pulsars with 400 MHz flux $>$ 20 mJy, and dispersion measure $<$ 150~pc
cm$^{-3}$, to increase the probability of detection as well as minimise 
possibility of scattering tails at the lower observing frequency. This left 
around 20 pulsars within the GMRT declination range which satisfied these 
conditions. In addition we also included PSR J2144--3933, which was the longest
period pulsar ($P$ = 8.510 seconds) known at the time of these observations. In
Table \ref{tab:obs} we report the details of 21 pulsars used in this work. The 
Table shows the widths \citep[$W_{10}$,][]{mit16} of the profiles, measured at 
10\% of the peak intensity level, which shows the on-pulse emission to be 
restricted to less than 20\% of the period in all cases. Table \ref{tab:obs} 
also lists the observing frequency, the period of the pulsar and the total 
phase-bins in the folded profile (Nbin), as well as the number of bins averaged
for the on-pulse (N$_{on}$) and off-pulse (N$_{off}$) images in each case.

An example of the folded profile from the interferometric observations of PSR 
B0628--28 is shown in Fig. \ref{fig:fold} which highlights the on-pulse and 
off-pulse regions. Detailed images corresponding to both the on-pulse and 
off-pulse windows were produced, and the region around each pulsar is shown as 
intensity contours (see Fig. \ref{fig:mapB0628}). The pulsar signal is affected
due to interstellar scintillation which causes quasi-periodic variations of 
flux within the observing duration. The imaging technique inherently assumes 
the flux of the sources to be constant for the duration of the observations. 
Any inherent flux variations causes phase errors around the source resulting in
increased noise levels. In some cases the scintillation was prominently present
as indicated by phase structures around the pulsar in the on-pulse images. The 
estimated average on-pulse flux ($S_{on}$) for the observing duration of each 
pulsar is shown in Table \ref{tab:obs}, along with the period averaged flux, 
$S_{avg}$ = $S_{on} N_{on} t_{res}/P$, where $t_{res}$ (= 0.128/0.256 seconds) 
is the time resolution of observations. In the Table the noise levels near the 
location of the pulsar is shown for both the on-pulse ($\sigma_{on}$) and 
off-pulse ($\sigma_{off}$) maps. The off-pulse noise varied between 60 $\mu$Jy 
and 260 $\mu$Jy, and was less than 100 $\mu$Jy for majority of pulsars. The 
on-pulse noise was higher due to the presence of the strong nearby pulsar which
also had phase errors due to scintillation.

\section{Results}\label{sec:result}

\begin{table*}
\caption{Analysis of Off-pulse Emission}
\begin{center}
\begin{tabular}{cccccccccccc}
\hline
PSR & $P$ & $W_{10}$ & Freq & Nbin & N$_{on}$ & N$_{off}$ & $S_{on}$ & $S_{avg}$ & $\sigma_{on}$ & $\sigma_{off}$ & $\frac{5\sigma_{off}}{S_{avg}}$ \\
\hline
   & (sec) & (\degr) & (MHz) &   &    &   &  (mJy)  &   (mJy)   &     (mJy)     &      (mJy)    &  (\%) \\
\hline
    &   &   &   &    &    &   &    &      &       &      &    \\
 B0138+59 & 1.223 & 30.2$\pm$0.8 & ~610 & ~9 & 3 & ~3 &  112.4$\pm$8.2~ & 34.7$\pm$2.5 & 0.154 & 0.078 & 1.12 \\
    &   &    &    &    &   &   &    &      &       &      &   \\
 B0148--06 & 1.465 & 37.9$\pm$0.1$^{\dagger}$ & ~610 & 11 & 3 & ~5 & 256.3$\pm$19.0 & 66.1$\pm$4.9 & 1.055 & 0.168 & 1.27 \\
    &   &    &    &    &    &   &    &      &       &      &  \\
 B0320+39 & 3.032 & 11.3$\pm$0.2 & 1280 & 23 & 3 & 14 & ~7.11$\pm$0.58 & 0.90$\pm$0.07 & 0.148 & 0.087 & 48.3 \\
    &   &    &    &    &    &   &    &      &       &      &  \\
 B0525+21 & 3.746 & 18.8$\pm$0.7 & 1280 & 29 & 5 & 17 & ~29.9$\pm$2.2~ & ~5.1$\pm$0.4~ & 0.154 & 0.077 & 7.55 \\
    &   &    &    &    &    &   &    &      &       &      &  \\
 B0628--28 & 1.244 & 37.4$\pm$0.2 & ~610 & ~9 & 3 & ~3 & 646.5$\pm$47.7 & 196.5$\pm$14.5 & 1.411 & 0.076 & 0.25$^*$ \\
    &   &    &    &    &    &   &    &      &       &      &  \\
 B0809+74 & 1.292 & 26.4$\pm$0.7 & ~610 & 10 & 3 & ~4 & ~93.2$\pm$6.9 & 27.3$\pm$2.0 & 0.207 & 0.060 & 1.10 \\
    &   &    &    &    &    &   &    &      &       &      &  \\
 B0818--13 & 1.238 & 10.1$\pm$0.2 & ~610 & ~9 & 3 & ~3 & 170.7$\pm$12.7 & 52.1$\pm$3.9 & 0.364 & 0.160 & 1.54 \\
    &   &    &    &    &    &    &   &      &       &      &  \\
 B0834+06 & 1.274 & 10.3$\pm$0.7 & 1280 & ~9 & 3 & ~3 & ~22.4$\pm$1.7~ & ~6.6$\pm$0.5~ & 0.228 & 0.165 & 12.50 \\
    &   &    &    &    &    &   &    &      &       &      &  \\
 B1237+25 & 1.383 & 13.7$\pm$0.4 & 1280 & 10 & 3 & ~4 & ~47.6$\pm$1.7 & ~12.9$\pm$0.5 & 0.128 & 0.081 & 3.14 \\
    &   &    &    &    &    &    &   &      &       &      &  \\
 B1738--08 & 2.043 & 18.0$\pm$0.1$^{\dagger}$ & ~610 & 15 & 3 & ~7 & 69.3$\pm$5.1 & 12.8$\pm$0.9 & 0.148 & 0.095 & 3.71 \\
    &   &    &    &    &    &   &    &      &       &      &  \\
 B1819--22 & 1.874 & 16.3$\pm$0.1$^{\dagger}$ & ~610 & 14 & 3 & ~7 & 90.5$\pm$6.6 & 18.3$\pm$1.3 & 0.198 & 0.148 & 4.04 \\
    &   &    &    &    &    &   &    &      &       &      &  \\
 B1839+56 & 1.653 & 12.1$\pm$0.7 & ~610 & 12 & 3 & ~5 & 71.7$\pm$5.3 & 16.4$\pm$1.2 & 0.254 & 0.127 & 3.87 \\
    &   &    &    &    &    &   &    &      &       &      &  \\
 B1846--06 & 1.451 & ~7.6$\pm$0.2 & ~610 & 11 & 3 & ~5 & 44.0$\pm$3.3 & 11.5$\pm$0.9 & 0.238 & 0.104 & 4.52 \\
    &   &    &    &    &    &   &    &      &       &      &  \\
 B1905+39 & 1.236 & 21.3$\pm$0.7 & ~610 & ~9 & 3 & ~3 & 22.1$\pm$1.7 & 6.8$\pm$0.5 & 0.237 & 0.202 & 14.85 \\
    &   &    &    &    &    &   &    &      &       &      &  \\
 B1919+21 & 1.337 & 11.8$\pm$0.2 & ~610 & 10 & 3 & ~4 & 182.3$\pm$13.5 & 51.5$\pm$3.8 & 0.583 & 0.260 & 2.52 \\
    &   &    &    &    &    &   &    &      &       &      &  \\
 B2045--16 & 1.961 & 16.4$\pm$0.7 & 1280 & 15 & 5 & ~7 & ~65.1$\pm$1.7 & 20.9$\pm$0.5 & 0.449 & 0.061 & 1.46 \\
    &   &    &    &    &    &   &    &      &       &      &  \\
 J2144--3933 & 8.510 & ~1.4$\pm$0.2$^{\dagger}$ & ~610 & 33$^{\ddagger}$ & 3 & 21 & 28.8$\pm$2.2 & ~2.6$\pm$0.2 & 0.420 & 0.090 & 17.31 \\
    &   &    &    &    &    &   &    &      &       &      &  \\
 B2154+40 & 1.525 & 27.0$\pm$0.8 & ~610 & 11 & 3 & ~5 & 104.2$\pm$7.7 & 25.8$\pm$1.9 & 0.142 & 0.088 & 1.71 \\
    &   &    &    &    &    &   &    &      &       &      &  \\
 B2303+30 & 1.576 & 8.6$\pm$0.7 & ~610 & 12 & 3 & ~5 & 51.0$\pm$3.7 & 12.2$\pm$0.9 & 0.113 & 0.076 & 3.11 \\
    &   &    &    &    &    &   &    &      &       &      &  \\
 B2319+60 & 2.256 & 25.8$\pm$0.8 & ~610 & 17 & 3 & 10 & 285.8$\pm$20.9 & 47.9$\pm$3.5 & 0.269 & 0.159 & 1.66 \\
    &   &    &    &    &    &   &    &      &       &      &  \\
 B2327--20 & 1.643 & ~7.4$\pm$0.1 & ~610 & 12 & 3 & ~5 & 40.0$\pm$2.9 & 9.2$\pm$0.7 & 0.142 & 0.080 & 4.35 \\
\hline
\hline
\end{tabular}
\\$^{\dagger}$Width measured at 5$\sigma$ of the baseline level; $^*$Using 
Off-pulse emission flux 0.5 mJy; $^{\ddagger}$256 millisecond integration.
\end{center}
\label{tab:obs}
\end{table*}

\begin{figure*}
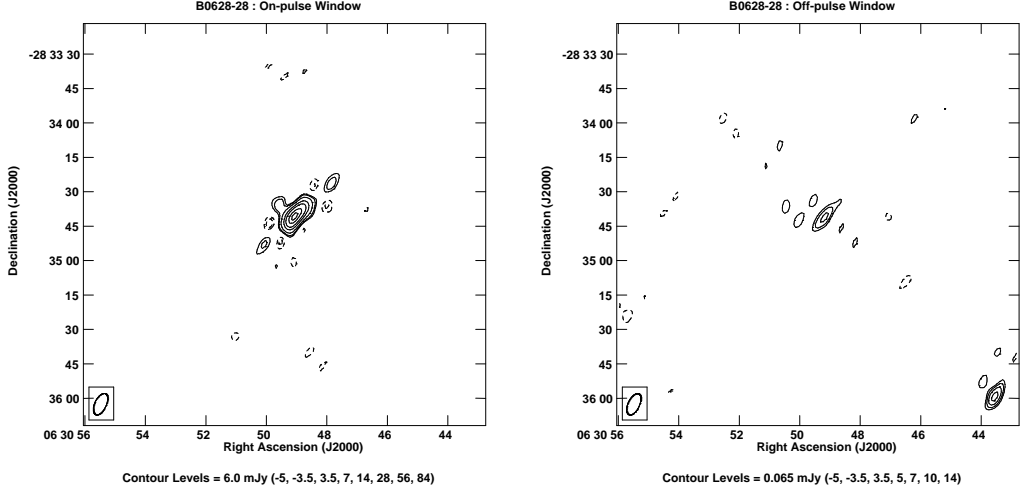

\centering
\begin{tabular}{@{}cr@{}}
{\mbox{\includegraphics[scale=0.35,angle=0.]{B0628N_cntr.ps}}} &
{\mbox{\includegraphics[scale=0.35,angle=0.]{B0628F_cntr.ps}}} \\
\end{tabular}
\caption{The figure shows the intensity contour map around the pulsar B0628--28
for the on-pulse (left panel) and off-pulse (right panel). The spatial 
resolution in the images is shown in the bottom left box. The on-pulse is seen 
as a point source in the center of the image with flux 646.5$\pm$47.7 mJy and 
noise levels $\sim$ 1-5 mJy around the source. The pulsar emission is affected 
by interstellar scintillation resulting in phase structures around it. A low 
level point source with peak flux 0.5 mJy is seen at the location of the pulsar
in the off-pulse image. The absence of the strong variable point source results
in significantly lower noise levels $\sim$ 60-80 $\mu$Jy around the central 
point source. An unrelated point source near the bottom right corner is also 
visible in the off-pulse image due to the lower noise levels.}
\label{fig:mapB0628}
\end{figure*}

\citet{bas11,bas12} reported detection of off-pulse emission from PSR 
B0525+21 and PSR B2045--16, however as discussed in section~\ref{sec:leak}, 
we found that this earlier detection was caused due to leakage from the 
on-pulse signal into the off-pulse window. We have corrected the source of 
leakage and searched for off-pulse emission in 21 pulsars. No clear point 
source structure was seen in the off-pulse images at the location of the pulsar
in most cases with the exception of  PSR B0630--28. Thus we confirm the results
of \citet{mar19} who has also reported the absence of off-pulse emission from 
PSR B0525+21 and PSR B2045--16. Their study used the European VLBI Network at 
1.39 GHz observing frequency. The noise levels in the off-pulse maps were 14 
$\mu$Jy for PSR B0525+21 and 32 $\mu$Jy in case of PSR B2045--16. These are 
significantly lower than the the noise levels obtained in this work at 1.28 
GHz, with 77 $\mu$Jy for PSR B0525+21 and 61 $\mu$Jy for PSR B2045--16. In case
of PSR B0525+21 the 5$\sigma$ detection limit at 1.39 GHz was 70 $\mu$Jy which 
was around 0.7\% of the average pulsar flux. In case of PSR B2045--16 the 
corresponding limit was 160 $\mu$Jy which was around 0.5\% of the average flux.
In addition we have also verified the non detection of off-pulse emission in 
these two pulsars at 325 MHz with noise levels around 200--500 $\mu$Jy and 610 
MHz where the noise levels were between 100--200 $\mu$Jy.

In PSR B0628--28 a weak point source structure was seen in the off-pulse window
with peak flux of 0.502 mJy, which is 6.5$\sigma$ of the noise level. The 
images corresponding to the on-pulse and off-pulse windows of PSR B0628--28 are
shown in Fig. \ref{fig:mapB0628}, where the presence of a point source in the 
off-pulse window coincident with the pulsar in the on-pulse image is seen. We 
have estimated the relative strength of off-pulse emission compared to the 
average flux ($S_{off}$/$S_{avg}$) which is around 0.25\% (see Table 
\ref{tab:obs}, last column). PSR B0628--28 is the brightest source on our 
sample with $S_{avg}$ = 196.5$\pm$14.5, which is between 3 to 80 times higher 
than the other pulsars observed at 610 MHz. The flux of the 5 pulsars observed 
at 1280 MHz is expected to be much lower compared to their 610 MHz values due 
to steep inverse power law nature of pulsar spectra \citep{mar00}. In Table 
\ref{tab:obs} (final column) the estimates of the ratio between the detection 
limit, defined as 5$\times\sigma_{off}$, and $S_{avg}$ is calculated. The 
minimum ratio is several times higher than 0.25\% which indicates that if the 
off-pulse emission is present in the other pulsars at a level similar to 
B0628--28, it will be well below the detection limit of the instrument. 

The low level off-pulse emission from PSR B0628--28 requires verification
from independent Telescope systems. Future studies of off-pulse emission in a 
significant number of pulsars will require higher sensitive observations with 
at least an order of magnitude lower levels of noise in the images. More 
significant detections of off-pulse emission, if present, will allow further 
detailed characterisation of their emission behaviour, like temporal 
variability, spectral nature, etc, which can be used to constrain the location 
of the emission.

\section{Discussion}\label{sec:disc}
\begin{figure}
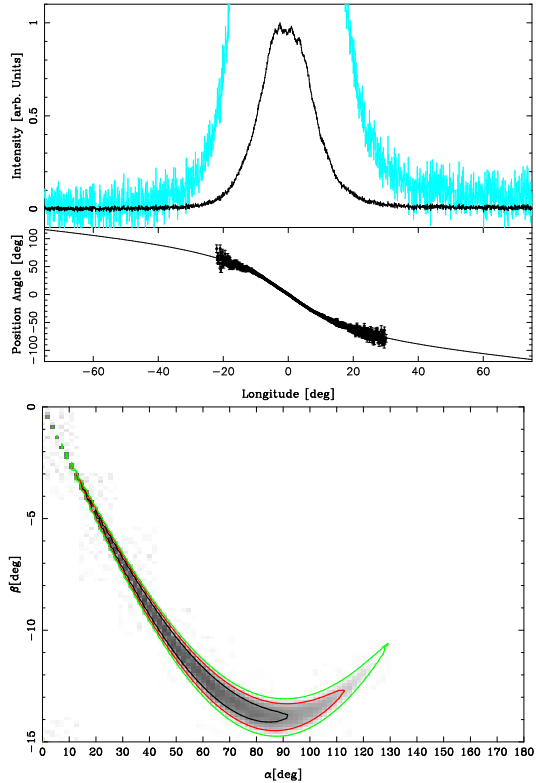

\centering
{\mbox{\includegraphics[scale=0.3,angle=-90.]{B0628_ppa.ps}}} \\
{\mbox{\includegraphics[scale=0.3,angle=-90.]{B0628_rvmchisq.ps}}} \\
\caption{The top panel shows the profile of PSR B0628--28 as well as the 
zoomed version (10 times the intensity, cyan) which highlights the low level 
emission near the profile edges. The lower window shows the variation of the 
polarization position angle (PPA) across the profile. The best approximation of
the rotating vector model to the PPA is also shown in the figure using geometry
specified by $\alpha$ = 13$\degr$ and $\beta$ = -3.1$\degr$. The bottom window 
shows the $\chi^2$ distribution for the fitted parameter $\alpha$ and $\beta$, 
where the contours in black red and green corresponds to 1, 2 and 3 times the 
minimum $\chi^2$ values. The parameters $\alpha$ and $\beta$ are highly 
correlated in the fit.}
\label{fig:ppaB0628}
\end{figure}

The off-pulse emission seen in pulsars can arise due to a variety of reasons 
which we explore below.

\subsection{Effect of Line of Sight Geometry}
The on-pulse emission corresponds to the region of open dipolar magnetic field 
line, and the duty cycle of the emission depends on the geometry of the pulsar.
n certain geometrical configurations, particularly for almost aligned rotator, 
there are instances where the pulsed emission can have almost 100\% duty cycle. 
In case of PSR B0628--28 the pulsed emission has a duty cycle of about 15\%. 
However, the possibility remains that due to some favourable alignment the 
observers' line of sight (LOS) cuts across emission beam over a wide longitude 
range, but the emission is at a significantly lower level in the majority of 
period and hence cannot be detected as a pulsed emission. In such a scenario 
the low level signal will be detected as off-pulse emission. The on-pulse 
emission height is constrained to be well within the pulsar magnetosphere at 
less than 10\% of the light cylinder radius \citep{kij98,mit02,kij03,mit04,
krz07,mit17}. Hence, in order to have the LOS to be within the emission beam 
for a large fraction of the pulsar period the magnetic inclination angle 
($\alpha$), i.e. the angle between the rotation and magnetic axis, has to be 
small. 

A possible way to resolve this issue is to understand the emission geometry of 
PSR B0628--28. \citet{ran90,ran93} developed the empirical theory of pulsar 
emission, where pulse width and polarization properties has been used to derive
the pulsar geometry. \citet{ran93} classified the pulsar PSR B0628--28 as a 
conal single and found $\alpha \sim 13.5^{\circ}$ and the angle between the 
rotation axis and the observer line of sight to be $\beta \sim 3.2^{\circ}$. To
verify the geometry at our observing frequency of 610 MHz we used the 
observations of PSR B0628--28 at 618 MHz obtained from the MSPES survey 
\citep{mit16}. Fig. \ref{fig:ppaB0628} shows the profile of PSR B0628--28 along
with the polarization position angle (PPA) across the profile. The PPA traverse
shows an S-shaped curve which is dependent on the geometrical angles $\alpha$ 
and $\beta$, as explained by the rotating vector model \citep[RVM,][]{rad69}. 
According to RVM the change in PPA ($\psi$) traverse reflects the change in 
projection of the magnetic field vector in the emission region as a function of
pulse rotational phase ($\phi$) and can be expressed as :
\begin{equation}
\psi = \tan^{-1} \left( \frac{\sin{\alpha}\sin{\phi}} {\sin{(\alpha+\beta)}\cos{\alpha} - \sin{\alpha}\cos{(\alpha+\beta)}\cos{\phi}}\right)
\label{eq_rvm}
\end{equation}

In Fig. \ref{fig:ppaB0628} the RVM fits to the PPA is shown which accurately 
reproduces the observed S-shaped curve \citep[also see][]{bec05}. However, a 
number of studies \citep{eve01,mit04} have shown that the RVM fits are not 
sufficient to estimate $\alpha$ and $\beta$ which are highly correlated in 
these fits as seen in the $\chi^2$ contour (see Fig. \ref{fig:ppaB0628}, 
bottom window). We found the best fit geometry as $\alpha=$ 13$\degr$
and $\beta=$ -3.1$\degr$, which is consistent with the estimates of 
\cite{ran93}. We used the smallest possible value of $\alpha$ which was within
the 1-$\sigma$ envelope of the $\chi^2$ distribution, to ensure maximum LOS
traverse of the emission beam. Note that our PPA fit is able to distinguish the
LOS to be an outer line of sight, i.e. $\beta$ having a negative value since 
the PPA traverse has a wider span and the characteristics flaring of the PPA in
the profile wings is evident. 

With the above estimate of $\alpha$ and $\beta$ and assuming the emission 
arises from constant height across the pulse, we can find the radius of the 
beam opening angle ($\rho$) using the expression :
\begin{equation}
\sin^2{\rho/2} = \sin{(\alpha+\beta)}\sin{\alpha}\sin^2{W/4}+\sin^2{\beta/2}
\label{eq_rho}
\end{equation}
where $W$ is the width of the pulse profile. Since our concern is to see how
far detectable pulse emission is present, we choose the profile edges at the
level of 5 times the off region rms and find $W = 51.3\degr$. Thus using 
$\alpha=$ 13$\degr$ and $\beta=$ -3.1$\degr$ we find $\rho = 5.8^{\circ}$. We 
thus confirm that the single component profile in this pulsar which has been 
classified as a conal single ($S_d$) type \citep{ran93}, is consistent with the
relatively shallow PPA traverse and comparatively high $\mid\beta/\rho\mid$ =
0.53. In a dipolar case the magnetic field lines diverge with emission height 
($h_{em}$), such that 
\begin{equation}
h_{em} = 10 \left(\rho/1.23\right)^2 (P/1~\mathrm{sec}) ~~ \mathrm{km},
\label{eq_hrho}
\end{equation}
where the beam opening angle at the stellar surface, with radius = 10 km, is 
equal to the radius of the opening angle of polar cap, $\rho_s$ = 1.23$\degr 
P^{-0.5}$. Using $\rho = 5.8^{\circ}$ and $P=1.24$ sec, we obtain the 
$h_{em} = 284$ km.  

\begin{figure*}
\centering
{\mbox{\includegraphics[scale=0.85,angle=0.]{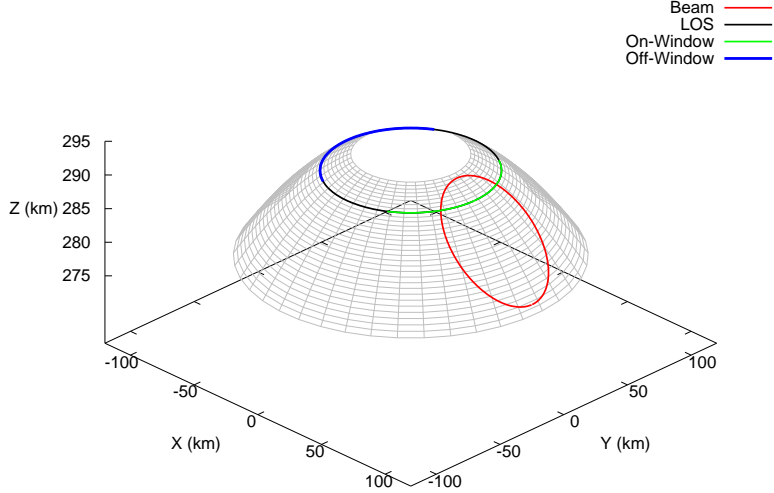}}} \\
\caption{The figure shows the line of sight (LOS) traverse across emission beam
in PSR B0628--28 at 618 MHz. The LOS regions corresponding to the on-pulse 
(green) and off-pulse (blue) windows used for the images are also shown in the
figure and indicates that the off-pulse window is well outside the emission 
beam. In the above configuration the dipolar magnetic axis at the center of the
emission beam is inclined at $\alpha=$ 13.0$\degr$ with the z-axis (rotation 
axis). The LOS has angular separation $\beta=$ -3.1$\degr$ from the magnetic 
axis, i.e $\alpha+\beta=$ 9.9$\degr$ with the z-axis as it rotates around it. 
The radius of the beam opening angle is $\rho$ = 5.8$\degr$.}
\label{fig:geomB0628}
\end{figure*}

The implications of the pulsar geometry on the off pulse emission is 
represented in figure \ref{fig:geomB0628} where we have shown the emission beam
at 618 MHz within the pulsar magnetosphere. The figure also shows the LOS 
traverse within the emission beam as specified by the estimated geometric 
angles. Although, the inclination angle between the rotation and magnetic axis 
is relatively low, it is still not sufficient for the LOS to be consistently 
within the emission beam throughout the rotation period. This is further 
highlighted in the figure, where we also show the part of the LOS traverse 
corresponding to the on and off-pulse windows. The only way by which regions of
the off-pulse window can be included in the open field line region is to
decrease $\alpha$ and/or increase $h_{em}$. The observed off-pulse emission is 
outside the on-pulse emission beam and thus is likely  to have a different 
physical origin.

\subsection{Diffuse Nebulae around B0628--28}\label{sec:pwn}
The low level off-pulse emission in PSR B0628--28 can also arise due to the 
presence of a diffuse nebula around the pulsar. \citet{bas11} presented simple 
arguments to show it is unlikely for an older long period pulsar to sustain an 
extended nebulae around it. A pulsar wind nebula (PWN) is generated when the 
relativistic wind from pulsars are confined by the surrounding medium resulting
in shock waves which are luminous across the electromagnetic spectrum in 
synchrotron, inverse Compton, and optical line emission from the shocked 
regions. In case of isolated pulsars like B0628--28 there are two possibilities
for the PWN to arise, either in the form of a `Static' PWN or `Bow-shock' PWN. 
The Static PWN corresponds to the case when the pulsar is at relative rest with
respect to the surrounding medium \citep{bla73} while the Bow-shock is usually 
seen when the pulsar velocity is faster than the velocity of the shock front 
\citep{gae00}. 

In case of Static PWN the radius of the shock front is given as $R_S = 
(\dot{E}/4\pi\rho_o)^{1/5} t^{3/5}$, where $\dot{E}$ is the spin-down energy 
loss, $t$ the pulsar age, $\rho_o=m_H n$, $m_H$ being proton mass and $n$ 
particle density of the ambient medium. The above expression can be used to 
estimate the required density of the surrounding interstellar medium (ISM) to 
harbour a static PWN which can be seen as off-pulse emission. The distance of 
PSR B0628--28 has been estimated to be around 0.3 kpc \citep{del09,yao17} and 
using the telescope resolution as $\sim5$", the upper limit for the size of the
possible unresolved nebula is around 0.006~pc. The density of the ambient 
medium using the size limit is given as 
$n=5.35\times10^{12}(\dot{E}_{32}t_6^3/R^5_{0.01})$ cm$^{-3}$ \citep{bas11}, 
where $\dot{E}_{32}$ is in units of 10$^{32}$ erg~s$^{-1}$, $t_6$ is in units 
of 10$^6$ years and $R_{0.01}$ is in 0.01~pc. Using $\dot{E}=1.5\times10^{32}$ 
erg~s$^{-1}$ and $t=2.77\times10^6$ year, the required ISM density for the 
Static PWN around B0628--28 is $\sim 10^{15}$ cm$^{-3}$. The typical densities
of ISM is around 0.03~cm$^{-3}$, which makes it highly improbable to find such
high density regions around the pulsar to sustain a Static PWN. 

The Bow-shock PWN is seen in young, highly energetic pulsars with 
$\dot{E}>10^{35}$ erg~s$^{-1}$ \citep{gae06}. These pulsars generally have high 
velocities $V_{PSR}>$ 500-1000 km/s, with respect to the surrounding medium 
which results in the formation of a Bow-shock instability with radius 
$R_{BS}\sim0.1-1$ pc. The pulsar B0628--28 is older, less energetic and has 
velocity of 77.29 km~s$^{-1}$ which is not suitable for the formation of a 
Bow-shock nebula. This is further highlighted by the estimates of the the 
radius of the Bow-shock which is given as $R_{BS}=(\dot{E}/4\pi c \rho_o 
V_{PSR}^2)^{0.5}$, where $c$ is the speed of light. The above expression can be
simplified as $R_{BS}=1.3\times10^{-3}(\dot{E}_{32}/n_{0.01}V_{100}^2)^{0.5}$ 
pc, where $V_{100}$ is in units of 100 km~s$^{-1}$ and $n_{0.01}$ is in units 
of 0.01~cm$^{-3}$. Using $V_{PSR}=$ 77.29 km~s$^{-1}$ and $n=$ 0.03~cm$^{-3}$, 
the estimated size of Bow-shock PWN around B0628--28 is 
$R_{BS}=1.2\times10^{-3}$ pc, showing that the radius of the possible 
bow shock is very small and hence unlikely to form a Bow-shock PWN.

\subsection{Cyclotron Resonance Instability in Outer Magnetosphere}\label{sec:cycres}
Another possible location of the off-pulse emission can be
the outer magnetosphere (closer to the light cylinder) along the pulsar open field lines. 
\citet{kaz87,kaz91} showed the 
possibility of cyclotron resonance instability to develop in the outflowing 
relativistic plasma near the outer magnetosphere, leading to coherent radio 
emission. This mechanism was considered to be a likely candidate for off-pulse 
emission by \citet{bas13}, where detailed calculations were carried out to 
explore the required plasma characteristics. The outflowing plasma along the 
open field lines of the pulsar magnetosphere plasma consists of an ultra 
relativistic beam of primary particles ($\gamma_b\sim10^6$) and a secondary 
cloud of electron positron pair plasma which is less energetic ($\gamma_s\sim 
10-1000$). The plasma is generated near the polar cap region of the pulsar and 
is constrained to move along the field lines due to the high value of magnetic 
field. However, the magnetic field becomes weaker near the outer magnetosphere 
($B_d \propto 1/r^3$), and the particles can gyrate and move across the field 
lines. Within the dense secondary plasma clouds a number of electromagnetic 
modes are generated, but if their amplitudes grow as they propagate through the
medium, they can escape the plasma as radiation. The cyclotron resonance 
instability is a likely mechanism where the extra-ordinary wave in the 
secondary pair plasma is amplified by the resonating primary particles gyrating
across the field lines. There are two quantities of interest, the growth rate 
($\Gamma$) of the plasma waves and the resonance frequency of the primary 
particles ($\omega_0$). The primary conditions for electromagnetic waves to be 
emitted are, significant increase in the wave amplitude, $\Gamma\tau\geq$1, 
where $\tau\sim P/2\pi$ is the typical growth time, and the resonance frequency
is less than the damping frequency in the medium, $\omega_0<\omega_1$ 
\citep{kaz91}.

\citet{bas13} have estimated the two conditions necessary for development of 
cyclotron resonance instability in the form of different pulsar parameters,
including period, period derivative ($\dot{P}$), the Lorentz factors of the 
different plasma particles, the screening factor ($\eta$) of the electric field
in the inner acceleration region above the polar cap, and the multiplicity 
factor ($\chi$). The growth factor, resonance frequency and damping frequency 
are given as :
\begin{eqnarray}
 \Gamma\tau&\approx&1.2\times10^{-15}\left({\displaystyle\frac{\eta\gamma_{res}^3}{\gamma_T}}\right)\left({\displaystyle\frac{P^3}{\dot{P}_{-15}}}\right)\nonumber \\
\omega_0&\approx&3.4\times10^{16}\left({\displaystyle\frac{\gamma_p^3}{\chi\gamma_{res}}}\right)\left({\displaystyle\frac{\dot{P}_{-15}}{P^4}}\right) \nonumber \\
\omega_1&\approx&6.5\times10^8~\gamma_p\left({\displaystyle\frac{\dot{P}_{-15}}{P^5}}\right)^{1/2}\nonumber
\label{eqn:cycres}
\end{eqnarray}
Typical values of the physical parameters in pulsars are $\gamma_{res}$ = 
2$\times$10$^{6}$ (primary resonant particles), $\gamma_T$ = 10$^2$ (the 
thermal spread in primary particle distribution), $\gamma_p$ = 10 (secondary 
plasma), $\chi$ = 10$^4$, and $\eta$ = 0.1. The pulsar B0628--28 has $P$ = 
1.244 seconds and $\dot{P}$ = 7.12$\times$10$^{-15}$ s/s. Using the above 
values the growth factor for cyclotron resonance instability in B0628--28 is 
$\Gamma\tau\approx2.5$, while the resonance frequency is 
$\nu_0=\omega_0/2\pi\approx800$ MHz and cutoff frequency is 
$\nu_1=\omega_1/2\pi\approx1.6$ GHz, respectively. This shows that the 
cyclotron resonance instability can develop in the outer magnetosphere of 
B0628--28 leading to coherent radio emission which can be seen as off-pulse 
emission. However a detailed characterization of the off-pulse emission at 
multiple radio frequencies are needed to establish a detailed model that
can explain the observed flux level of the off-pulse emission.

\section{Summary}
We report on a search for off-pulse emission from 21 long period pulsars using 
observations from the GMRT. Off-pulse emission was earlier reported in PSR 
B0525+21 and PSR B2045--16 by \citet{bas11,bas12}. In a subsequent study using 
the EVN, \citet{mar19} showed the absence of off-pulse emission from these two 
pulsars. We have uncovered that the earlier detection of off-pulse emission was
a result of leakage from the on-pulse to the off-pulse window, and thereby 
confirm the results of \citet{mar19}. Low level off-pulse emission, with peak 
flux of $\sim$ 0.5 mJy, was detected in PSR B0628--28. More sensitive 
observations are required to confirm this detection and further characterise 
the off-pulse emission behaviour. The estimates of line of sight geometry makes
it unlikely for the off-pulse emission to be a low level emission feature 
within the main pulse emission beam. The presence of diffuse wind nebulae 
around the pulsar resulting in the observed off-pulse emission is also 
unlikely. On the other hand it is possible for the cyclotron resonance 
instability to develop in the outer magnetosphere of PSR B0628--28, which is a 
likely candidate for off-pulse emission.

\section*{Acknowledgments}
We thank the referee for the comments which helped to improve the paper. We 
thank the staff of the GMRT who have made these observations possible. The GMRT
is run by the National Centre for Radio Astrophysics of the Tata Institute of 
Fundamental Research. DM acknowledge the support of the Department of Atomic 
Energy, Government of India, under project no. 12-R\&D-TFR-5.02-0700. DM 
acknowledges funding from the grant ``Indo-French Centre for the Promotion of 
Advanced Research - CEFIPRA" grant IFC/F5904-B/2018.


\begin{thebibliography}{99}
\bibitem[Bartel \etal\ (1985)]{bar85} Bartel, N.; Ratner, M.I.; Shapiro, I.I.; Cappallo, R.J.; Rogers, A.E.E.; Whitney, A.R.  1985, AJ, 90, 318
\bibitem[Basu \etal\ (2011)]{bas11} Basu, R.; Athreya, R.; Mitra, D.  2011, ApJ, 728, 157 
\bibitem[Basu \etal\ (2012)]{bas12} Basu, R.; Mitra, D.; Athreya, R.  2012, ApJ, 758, 91 
\bibitem[Basu \etal\ (2013)]{bas13} Basu, R.; Mitra, D.; Melikidze, G.I.  2013, ApJ, 772, 86 
\bibitem[Basu \etal\ (2015)]{bas15} Basu, R.; Mitra, D.; Rankin, J.  2015, ApJ, 798, 105 
\bibitem[Becker \etal\ (2005)]{bec05} Becker, W.; Jessner, A.; Kramer, M.; Testa, V.; Howaldt, C.  2005, ApJ, 633, 367
\bibitem[Blandford \etal\ (1973)]{bla73} Blandford, R.D.; Ostriker, J.P.; Pacini, F.; Rees, M.J.  1973, A\&A, 23, 145 
\bibitem[Blaskiewicz \etal\ (1991)]{bla91} Blaskiewicz M.; Cordes J.M.; Wasserman I.  1991, ApJ, 370, 643 
\bibitem[Darling(2020a)]{dar20a} Darling, J.  2020a, PhysRevLett, 125, 121103
\bibitem[Darling(2020b)]{dar20b} Darling, J.  2020b, ApJ, 900L, 28 
\bibitem[Deller \etal\ (2009)]{del09} Deller, A.T.; Tingay, S.J.; Bailes, M.; {\etal} 2009, \apj, 701, 1243
\bibitem[Dyks(2008)]{dyk08} Dyks 2008, MNRAS, 391, 859
\bibitem[Everett \& Weisberg(2001)]{eve01} Everett J.E., Weisberg J.M., 2001, \apj, 553, 341
\bibitem[Dzib \etal\ (2018)]{dzi18} Dzib, S.A.; Rodríguez, L.F.; Karuppusamy, R.; Loinard, L.; Medina, S.X.  2018, ApJ, 866, 100
\bibitem[Gaensler \etal\ (1998)]{gae98} Gaensler, B.M.; Stappers, B.W.; Frail, D.A.; Johnston, S.  1998, ApJ, 499, 69
\bibitem[Gaensler \etal\ (2000)]{gae00} Gaensler, B.M.; Stappers, B.W.; Frail, D.A.; Moffett, D.A.; Johnston, S.; Chatterjee, S. 2000, MNRAS, 318, 58
\bibitem[Gaensler \& Slane(2006)]{gae06} Gaensler, B.M.; Slane, P.O.  2006, ARA\&A, 44, 17
\bibitem[Hankins \etal\ (1993)]{han93} Hankins, T.H.; Moffett, D.A.; Novikov, A.; Popov, M.  1993, ApJ, 417, 735
\bibitem[Kazbegi \etal\ (1987)]{kaz87} Kazbegi, A.Z.; Machabeli, G.Z.; Melikidze, G.I.  1987, AuJPh, 40, 755
\bibitem[Kazbegi \etal\ (1991)]{kaz91} Kazbegi, A.Z.; Machabeli, G.Z.; Melikidze, G.I.  1991, MNRAS, 253, 377
\bibitem[Kijak \& Gil(1998)]{kij98} Kijak, J.; Gil, J.  1998, MNRAS, 299, 855
\bibitem[Kijak \& Gil(2003)]{kij03} Kijak, J.; Gil, J.  2003, A\&A, 397, 969
\bibitem[Krzeszowski \etal\ (2007)]{krz07} Krzeszowski, K.; Mitra, D.; Gupta, Y.; Kijak, J.; Gil, J.; Acharyya, A.  2007, 393, 1617
\bibitem[Marcote \etal\ (2019)]{mar19} Marcote, B.; Maan, Y.; Paragi, Z.; Keimpema, A.  2019, A\&A, 627, L2 
\bibitem[Maron \etal\ (2000)]{mar00} Maron, O.; Kijak, J.; Kramer, M.; Wielebinski, R.  2000, A\&AS, 147, 195 
\bibitem[Mitra \& Rankin(2002)]{mit02} Mitra, D.; Rankin, J.  2002, ApJ, 577, 322
\bibitem[Mitra \& Li(2004)]{mit04} Mitra, D.; Li, X.H.  2004, A\&A, 421, 215
\bibitem[Mitra \etal\ (2016)]{mit16} Mitra, D.; Basu, R.; Maciesiak, K.; Skrzypczak, A.; Melikidze, G.I.; Szary, A.; Krzeszowski, K.  2016, ApJ, 833, 28
\bibitem[Mitra(2017)]{mit17} Mitra, D.  2017, JApA, 38, 52
\bibitem[Perry \& Lyne(1985)]{per85} Perry, T.E.; Lyne, A.G. 1985, MNRAS, 212, 489
\bibitem[P{\'e}tri \& Mitra(2020)]{pet20} P{\'e}tri J.; Mitra D. 2020, MNRAS, 491, 80
\bibitem[Radhakrishnan \& Cooke(1969)]{rad69} Radhakrishnan V., Cooke D.J.,  1969, Astrophys. Lett., 3, 225
\bibitem[Rankin(1990)]{ran90} Rankin, J.M.  1990, ApJ, 352, 247
\bibitem[Rankin(1993)]{ran93} Rankin, J.M.  1993, ApJ, 405, 285
\bibitem[Ravi \& Deshpande(2018)]{rav18} Ravi, K.; Deshpande, A.A.  2018, ApJ, 859, 22
\bibitem[Roy \etal\ (2010)]{roy10} Roy, J.; Gupta, Y.; Pen, U.L.; Peterson, J.B.; Kudale, S.; Kodilkar, J.  2010, ExA, 28, 25
\bibitem[Ruan \etal\ (2020)]{rua20} Ruan, D.; Taylor, G.B.; Dowell, J.; Stovall, K.; Schinzel, F.K.; Demorest, P.B.  2020, MNRAS, 495, 2125
\bibitem[Stappers \etal\ (1999)]{sta99} Stappers, B.W.; Gaensler, B.M.; Johnston, S.  1999, MNRAS, 308, 609
\bibitem[Strom \& Van Someren Greve(1990)]{str90} Strom, R.G.; Van Someren Greve, H.W.  1990, Ap\&SS, 171, 351
\bibitem[Swarup \etal\ (1991)]{swa91} Swarup, G.; Ananthakrishnan, S.; Kapahi, V.K.; {\etal}  1991, CuSc, 60, 95
\bibitem[Weiler \etal\ (1974)]{wei74} Weiler, K.W.; Goss, W.M.; Schwarz, U.J.  1974, A\&A, 35, 473
\bibitem[Yao \etal\ (2017)]{yao17} Yao, J.M.; Manchester, R.N.; Wang, N.  2017, ApJ, 835, 29

\end{thebibliography}
\end{document}